\documentclass[aps,prl,twocolumn,superscriptaddress,showpacs]{revtex4-1}
\usepackage{amsmath}
\usepackage{braket}
\usepackage{graphicx}
\usepackage{hyperref}

\begin{document}
    \title{Detection of Spin Entanglement via Spin-Charge Separation in Crossed
        Tomonaga-Luttinger Liquids}

    \author{Alexander Schroer}
    \affiliation{Institut f\"ur Mathematische Physik, Technische Universit\"at
        Braunschweig, D-38106 Braunschweig, Germany}
    \author{Bernd Braunecker}
    \affiliation{Scottish Universities Physics Alliance, School of Physics and 
        Astronomy, University of St Andrews, North Haugh, St Andrews KY16 9SS, 
        United Kingdom}
    \author{Alfredo Levy Yeyati}
    \affiliation{Departamento de F\'isica Te\'orica de la Materia Condensada,
        Condensed Matter Physics Center (IFIMAC), \\ and Instituto Nicol\'as
        Cabrera, Universidad Aut\'onoma de Madrid, E-28049 Madrid, Spain}
    \author{Patrik Recher}
    \affiliation{Institut f\"ur Mathematische Physik, Technische Universit\"at
        Braunschweig, D-38106 Braunschweig, Germany}
    \affiliation{\hbox{Interactive Research Center of Science, Tokyo Institute 
        of Technology, 2-12-1 Ookayama, Meguro, Tokyo 152-8551, Japan}}

    \begin{abstract}
        We investigate tunneling between two spinful Tomonaga-Luttinger liquids
        (TLLs) realized, e.g., as two crossed nanowires or quantum Hall edge
        states. When injecting into each TLL one electron of opposite spin,
        the dc current measured after the crossing differs for singlet, triplet,
        or product states. This is a striking new non-Fermi liquid feature
        because the (mean) current in a noninteracting beam splitter is
        insensitive to spin entanglement. It can be understood in terms of
        collective excitations subject to spin-charge separation.  This behavior
        may offer an easier alternative to traditional entanglement detection
        schemes based on current noise, which we show to be suppressed by the
        interactions.
    \end{abstract}


    \maketitle

    Entanglement is a necessary prerequisite for universal quantum computation
    and certain quantum communication protocols like quantum teleportation or
    dense coding \cite{nielsen00}. The creation of nonlocal pairwise entangled
    particles has been successfully demonstrated with photons \cite{freedman72,
    aspect82, weihs98} by violating a Bell inequality \cite{bell64, clauser69}.
    The same has not yet been demonstrated in transport experiments in a solid
    state device.  In particular, spin-entangled electrons are important
    candidates, because the electron spin in quantum dots could be used as a
    qubit \cite{loss98}, with proven promising spin-coherence
    times \cite{hanson07}. From the theoretical side, Cooper-pair splitters
    (CPSs)
    \cite{recher01,lesovik01,recher02,bena02,recher03,yeyati07,
    cayssol08,sato10} were proposed as a potential source of mobile and
    nonlocal spin-entangled pairs, using the process of crossed Andreev
    reflection \cite{torres99,falci01}. Experimentally, such CPSs have been
    built successfully \cite{hofstetter09, herrmann10, das12} with high
    efficiency \cite{schindele12}. However, the spin entanglement of these
    correlated pairs has not been demonstrated so far. Several detection schemes
    for entangled states were proposed based on a violation of a Bell inequality
    using cross-correlation (noise) measurements \cite{kawabata01,
    chtchelkatchev02, samuelsson03,beenakker03,samuelsson04-1,sauret05,chen12},
    current measurements in a CPS with spin-filter properties
    \cite{braunecker13}, or exploiting beam splitters
    \cite{burkard00,egues02,burkard03,hu04,
    samuelsson04-2,egues05,giovannetti06,prada06,mazza13} where a bunching or an
    antibunching behavior in the two-electron scattering process depends on the
    orbital wave function of the entangled pairs, distinguishing singlets from
    triplets or product states. The latter is an effect of statistics and holds
    already for noninteracting electrons. The average current does not carry a
    signature of entanglement in Fermi-liquid systems \cite{burkard00}.
    
    In this Letter we show that the situation is radically different in the case
    of a beam splitter made of one-dimensional interacting nanowires
    \cite{komnik98,komnik01}, or, almost equivalently, integer quantum Hall (QH)
    edge states \cite{ji03,neder06,neder07,bocquillon13-1,wahl14}. In these
    systems, which can conveniently be described as Tomonaga-Luttinger liquids
    (TLLs) \cite{tomonaga50,luttinger63}, the average current \emph{is}
    sensitive to spin entanglement due to the property of spin-charge
    separation.  This is a desirable feature because the current is generally
    much easier to measure than noise or higher-order correlation functions. An
    interpretation of the current noise in terms of (anti-)bunching
    \cite{burkard00} still applies, although Coulomb repulsion reduces the
    signal. The TLL system allows for an entangler \cite{recher02,bena02,sato10}
    and detector scheme without the need of magnetic elements (spin filters) nor
    noise-correlation measurements.  Experimentally, transport through crossed
    1-D conductors has already been demonstrated, including TLL effects
    \cite{kim01,gao04,karmakar11,bocquillon13-2}. We focus on the slightly more
    general case of nanowires and give details about a QH implementation in the
    Supplemental Material.

\begin{figure}
        \includegraphics{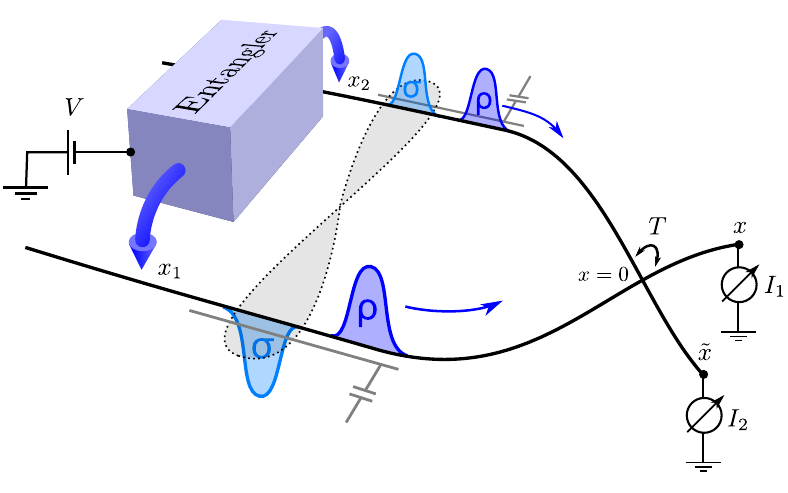}
        \vspace{-.1cm}
        \caption{(Color online) Tunnel junction with amplitude $T$ at $x=0$
            between two interacting one-dimensional wires. Via an entangler
            biased with a voltage $V$, two spin-entangled electrons are injected
            simultaneously at $x_1$ in wire~1 and at $x_2$ in wire~2 with an
            amplitude $I$, and subsequently decay into collective spin and
            charge excitations. The current expectation values $I_{1,2}$
            measured at $x,\tilde x$ at the far opposite side of the junction
            and their cross correlations are influenced by the entanglement of
            the original electrons.}
        \label{fig-setup}
    \end{figure}

    \emph{Model.}---%
    We consider two long nanowires (wire~1 and wire~2), which are connected
    through a weak tunnel junction at $x=0$ (Fig.~\ref{fig-setup}). To the left
    of the junction, at $x_1,x_2<0$, electrons are injected pairwise from an
    entangler, biased with a voltage $V$. The temperature is assumed smaller
    than the bias voltage and can be set to zero for convenience. The rate of
    injection is sufficiently low that there are no correlations between
    subsequent electron pairs.

    In the TLL left- and right-moving electron modes are expressed as bosonic
    fluctuations described by the Hamiltonian \cite{delft98,grabert01,
    giamarchi03}
    \begin{equation}
        H_0=\sum_{j\alpha}\int dx
            \frac{\hbar v_\alpha}{2}
            \Bigl[g_\alpha\Bigl
                    (\frac{\partial\phi_{j\alpha}}{\partial x}\Bigr)^2
                +\frac{1}{g_\alpha}\Bigl(\frac{\partial\theta_{j\alpha}}
                    {\partial x}\Bigr)^2\Bigr],
        \label{eq-h0}
    \end{equation}
    with $\phi$ and $\theta$ dual phase fields obeying
    $[\theta_{j\alpha}(x,t),\phi_{j'\alpha'}(x',t)]=(i/2)
    \delta_{jj'}\delta_{\alpha\alpha'}\text{sgn}(x-x')$. In this notation,
    $\partial_x\phi_{j\alpha}$ is proportional to the charge current
    ($\alpha=\rho$) or the spin current ($\alpha=\sigma$) in wire $j\in\{1,2\}$
    and $\partial_x\theta_{j\alpha}$ to the corresponding density. Assuming 
    SU(2) spin invariance, the interaction parameter in the spin sector is 
    $g_\sigma=1$, and in the charge sector, $g_\rho\equiv g<1$.
    With $v_F$ the Fermi velocity, $v_\rho=v_F/g_\rho$ and
    $v_\sigma=v_F$ are the velocities of spin and charge excitations. 
    The physical electron field with spin $s\in\{\uparrow,\downarrow\}$ is
    $\psi_{js}=\psi_{jRs}+ \psi_{jLs}$ with the
    right($R$)- and left($L$)-moving contribution
    $\psi_{jR/Ls}(x,t)=(2\pi a)^{-1/2}F_{jR/Ls} \exp[\pm i
    k_Fx+2\pi i\Phi_{jR/Ls}(x,t)]$ and $\Phi$ a linear combination
    of $\phi$ and $\theta$ \cite{giamarchi03,grabert01}. The Klein factors
    $F$ are unitary, anticommuting operators ensuring fermionic commutation
    relations \cite{grabert01}. The cutoff parameter of the wires, $a$,
    corresponds to their inverse bandwidth.

    The tunnel junction at $x=0$ is described by the Hamiltonian
    \begin{align}
        H_T&=T\hspace{-.5cm}\sum_{\substack{
                \nu,\nu'\in\{R,L\}\\s\in\{\uparrow,\downarrow\}}}
            \hspace{-.5cm}\Bigl(
            \psi_{1\nu's}^\dagger(0)\psi_{2\nu s}(0)
            +\psi_{2\nu's}^\dagger(0)\psi_{1\nu s}(0)
            \Bigr).
        \label{eq-ht}
    \end{align}

    \emph{Initial state approach.}---%
    First, we include the entangled electron pair
    as a suitably chosen \emph{initial state} at time $t_0$ with one electron in
    each wire at $x_1$ and $x_2$, respectively, on top of the many-particle
    ground state $\Ket{}$
    \begin{align}
        \Ket{\varphi}&=\frac{\pi a}{\sqrt{2}}\Bigl(
            \psi^\dagger_{2\downarrow}(x_2)
            \psi^\dagger_{1\uparrow}(x_1)
            +e^{i\varphi}
            \psi^\dagger_{2\uparrow}(x_2)
            \psi^\dagger_{1\downarrow}(x_1)
            \Bigr)\Ket{}
            \notag\\
        &:=2^{-1/2}
            \sum_{\nu_1,\nu_2}\Bigl(
            \Ket{\nu_1\uparrow,\nu_2\downarrow}
            +e^{i\varphi}
            \Ket{\nu_1\downarrow,\nu_2\uparrow}\Bigr).
        \label{eq-expvalue}
    \end{align}
    The relative phase $\varphi$ is the rotation angle between the pure triplet
    state ($\varphi=0$) and the pure singlet state ($\varphi=\pi$). We choose
    $|x_{1,2}|\gg a$ to avoid initial overlap between the injected electrons and
    the tunnel contact. Later, we will show that the results of this model carry
    over to the case of an applied bias voltage $V$ by essentially replacing the
    wave-packet width $a$ of the state $\Ket{\varphi}$ by $\hbar v_F/eV$, where
    $e$ is the electron charge.  

    Every expectation value of an operator $\mathcal{O}$ with respect to these
    states can be written as $\bra{\varphi}\mathcal{O}\ket{\varphi}=
    \mathcal{O}^\text{dir}+\cos(\varphi)\,\mathcal{O}^\text{exc}$, where the
    \emph{direct} term $\mathcal{O}^\text{dir}=\sum_{\nu_1\nu_2}
    \bra{\nu_1\uparrow,\nu_2\downarrow}\mathcal{O}
    \ket{\nu_1\uparrow,\nu_2\downarrow}$ is the
    product state contribution, and the \emph{exchange} term
    $\mathcal{O}^\text{exc}=\sum_{\nu_1\nu_2}\bra{\nu_1\uparrow,\nu_2\downarrow}
    \mathcal{O}\ket{\nu_1\downarrow,\nu_2\uparrow}$ is a distinctive indicator
    of entanglement \cite{burkard00}. Varying $\varphi$ is a powerful way to
    identify the exchange contribution in a measurement, which will be discussed
    later on.

    Within this approach, the current expectation value in wire~1 after the
    injection is given by
    \begin{equation}
        I_1=e\Gamma_{2e}\int_{t_0}^{\infty}dt
            \Bra{\varphi}I_1(x,t)\Ket{\varphi},
        \label{eq-i1}
    \end{equation}
    with $x\gg a$, $\Gamma_{2e}\ll v_F a^{-1}$ the rate of injection, and
    the bosonized current operator \cite{giamarchi03}
    $I_j(x,t)=-\sqrt{2/\pi}\partial_t\theta_j(x,t)$. Similarly, the
    zero-frequency cross-correlations between the two wires are
    \begin{equation}
        S_{12}=\frac{e^2\Gamma_{2e}}{2}\int_{t_0}^{\infty}
            dt dt'\Bra{\varphi}
            \Bigl\{\delta I_1(x,t),
            \delta I_2(\tilde x,t')\Bigr\}
            \Ket{\varphi},
        \label{eq-s12}
    \end{equation}
    where $\delta I_j=I_j-\Bra{\varphi}I_j\Ket{\varphi}$.

    Treating $H_{\rm T}$ as a perturbation \footnote{In a spinless TLL at
    sufficiently strong interactions, $g<1/2$, electrostatic interwire effects
    become relevant (in the renormalization group sense), which change the model
    substantially \cite{komnik98}. Additional noninteracting channels like spin
    lower this bound \cite{gao04}, but we shall conservatively limit this
    discussion to moderate interaction strengths $1/2<g\le1$.}, the expressions
    Eqs.~\eqref{eq-i1} and \eqref{eq-s12} can be evaluated with a standard
    Keldysh nonequilibrium generating functional approach
    \cite{kane94,crepieux03,dolcini05}. Besides the zeroth-order contributions
    (no tunnel processes) $I_1^{(0)}=-e\frac{\Gamma_{2e}}{2}$ and
    $S_{12}^{(0)}=0$ they yield second order in $T$ \emph{direct} and
    \emph{exchange} corrections. The former contain effects due to interactions
    and spin-charge separation, which are further discussed in the Supplemental
    Material, but they are not sensitive to entanglement. The latter are
    \begin{align}
        \label{eq-i2exc}
        I_1^{(2)\text{exc}}&=e\Gamma_{2e}
            \frac{1+g}{2} \\
            \times&\Bigl[
            \Bra{R\uparrow,R\downarrow}
                U^{(1)\uparrow\dagger}_{1R\rightarrow 2R}
                U^{(1)\downarrow}_{1R\rightarrow 2R}
                \Ket{R\downarrow,R\uparrow} \notag\\
            -&\Bra{R\uparrow,R\downarrow}
                U^{(1)\downarrow\dagger}_{2R\rightarrow 1R}
                U^{(1)\uparrow}_{2R\rightarrow 1R}
                \Ket{R\downarrow,R\uparrow}
            \Bigr], \notag
    \end{align}
    \begin{align}
        \label{eq-s2exc}
        S_{12}^{(2)\text{exc}}&=-e^2\Gamma_{2e}
            \Bigl(\frac{1+g}{2}\Bigr)^2 \\
            \times\text{Re}\Bigl[
            &\Bra{R\uparrow,R\downarrow}
                U^{(1)\uparrow\dagger}_{1R\rightarrow 2R}
                U^{(1)\downarrow}_{1R\rightarrow 2R}
                \Ket{R\downarrow,R\uparrow} \notag\\
            +&\Bra{R\uparrow,R\downarrow}
                U^{(1)\downarrow\dagger}_{2R\rightarrow 1R}
                U^{(1)\uparrow}_{2R\rightarrow 1R}
                \Ket{R\downarrow,R\uparrow}
            \Bigr]. \notag
    \end{align}

    Here, $U^{(1)s}_{jR\rightarrow kR}=-i\hbar^{-1}\int_{t_0}^{\infty}
    dt'\;H_T(t')|^s_{jR\rightarrow kR}$ is the
    first-order contribution of the time evolution operator which connects the
    initial state to a final state in the distant future, including only the
    parts of the tunnel Hamiltonian $H_T$ which describe tunneling
    of right-moving spin $s$ electrons from wire $j$ into wire $k$. In
    this way we can distinguish two events: an electron tunnels out of wire~1
    ($1\rightarrow2$), and an electron tunnels into wire~1 ($2\rightarrow1$).
    One increases and the other decreases the current, but both add to the
    noise. Their strength is given by the overlap of the corresponding final
    state $U^{(1)s}\Ket{\downarrow\uparrow}$ with its spin-flipped counterpart
    $\Bra{\uparrow\downarrow}U^{(1)-s\dagger}$; i.e., a process has a large rate
    if the final state after one tunnel event is mostly invariant under
    spin flip. This will be a key observation to interpret the results. The
    factor $\frac{1+g}{2}$ is caused by charge fractionalization \cite{pham00}.
    When measuring the current or noise not in the TLL, but in Fermi liquid
    reservoirs to the right of the beam splitter, the complete charge will be
    detected \cite{maslov95,ponomarenko95,safi95,crepieux03,
    dolcini05,lebedev05, recher06}. Formally, this corresponds to setting
    $g\rightarrow 1$ in the prefactors (but not in the correlation functions) of
    Eqs.~(\ref{eq-i2exc}) and (\ref{eq-s2exc}). This, however, only leads to
    minor quantitative changes, so we will not make the distinction in the
    following. 

    In the noninteracting ($g=1$) and in the symmetric ($x_1=x_2$) cases, the
    time integrals in Eqs.~\eqref{eq-i2exc} and \eqref{eq-s2exc} can be solved
    analytically. At $g=1$, the exchange noise is a Lorentzian of $d=x_2-x_1$,
    \begin{align}
        S_{12}^{(2)\text{exc}}&=e^2\frac{\Gamma_{2e}}{2}
            \Bigl|\frac{T}{\hbar v_F}\Bigr|^2
            \frac{1}{4+(d/a)^2},
        \label{eq-s2exc-d0}
    \end{align}
    meaning that there can only be an exchange process if the spins meet at the
    tunnel junction. Like in earlier noninteracting results in energy
    \cite{burkard00,hu04,samuelsson04-2} and time domain \cite{olkhovskaya08},
    nonzero exchange noise requires orbital overlap. Interactions decrease the
    exchange signal. At $x_1=x_2$ the power law $S_{12}^{(2)\text{exc}}\propto
    ((1+g)/2)^2 (2g^{-1}+1)^{-(g^{-1}+g)/2}$ is obtained.
    
    \begin{figure}
        \includegraphics[width=9cm]{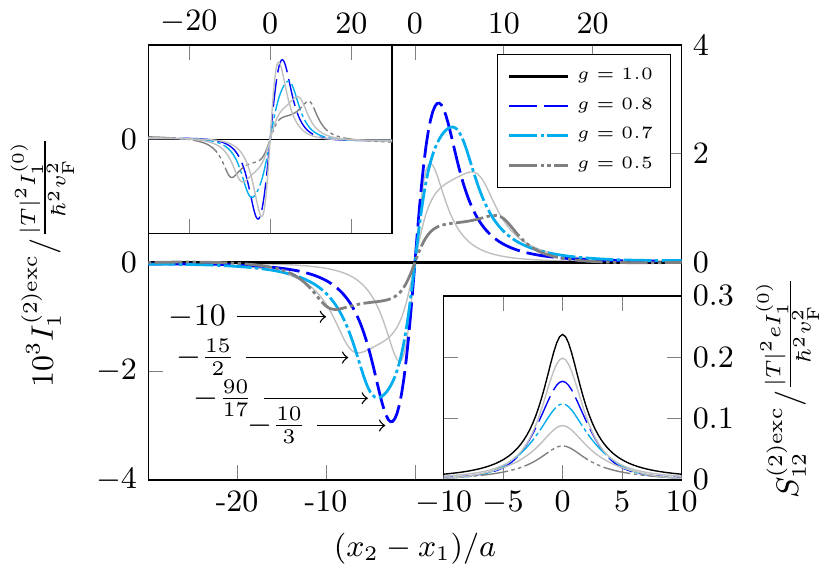}
        \caption{(Color online) Exchange contributions to the tunnel current in
            wire~1 and the zero-frequency current cross-correlations between
            wire~1 and wire~2 (right inset) for different interaction parameters
            $g$ and injection distances. $(x_1+x_2)/2=-15a$ is fixed. The
            exchange contribution to the tunnel current is nonzero if
            $x_1\approx x_2$ because spin-charge separation induces an asymmetry
            between the two directions of tunneling. The arrow tips indicate the
            expected positions of the maxima, cf. Eq.~\eqref{eq-opti}. Gray
            lines represent equidistant intermediate $g$ values. The exchange
            part of the current noise is finite only if the spins meet at the
            junction.  Left inset: analytic approximation,
            Eq.~\eqref{eq-estimate}.}
        \label{fig-initial}
    \end{figure}

    As expected, the exchange current vanishes exactly without interactions,
    since the two amplitudes in Eq.~(\ref{eq-i2exc}) cancel. This is already
    true when spin-charge separation is neglected, i.e., when setting
    $v_\rho=v_\sigma$.  For $v_\rho\neq v_\sigma$, however, a numerical
    integration of Eq.~\eqref{eq-i2exc} demonstrates that $I_1^{(2)\text{exc}}$
    is nonzero in general (Fig.~\ref{fig-initial}). This confirms that
    entanglement can be detected in the many-body system by current measurements
    only and that the phenomenon of spin-charge separation is essential. It
    induces a crucial asymmetry between the two competing processes, which goes
    unnoticed if both are summed up (current noise), but is relevant if they are
    subtracted (mean current).

\begin{figure}
        \includegraphics{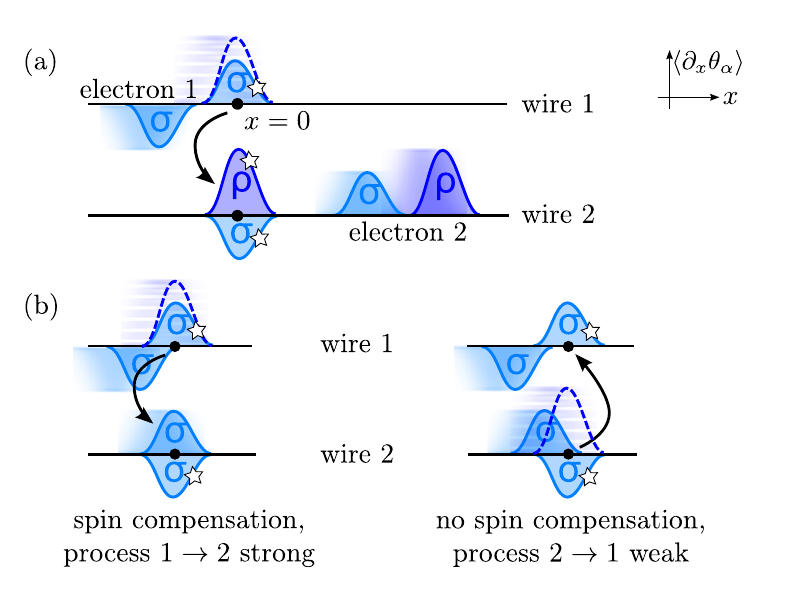}
        \caption{(Color online) Exchange process. (a) $x_1\ll x_2$. When the 
            charge excitation of electron~1 (dashed line) reaches the tunnel
            junction at $x=0$, the charge imbalance can trigger a tunnel event.
            This creates a new charge and a new spin excitation in wire~2 and
            leaves behind a spin hole in wire~1 (all marked by stars). Spin and
            charge excitations are drawn with different height for better
            visibility. (b) For suitable injection points
            $x_1/v_\rho=x_2/v_\sigma$ the new spin excitations compensate the
            one already present in each wire, leading to a strong exchange
            process. The competing process cannot have spin compensation at the
            same time and is weak. This asymmetry caused by spin-charge
            separation gives rise to a finite exchange current.}
        \label{fig-interpretation}
    \end{figure}

    The behavior of the exchange current in Fig.~\ref{fig-initial} can be
    qualitatively understood in the following way: In the nanowires, the two
    injected electrons decay each into a collective charge density excitation
    $\Bra{R\uparrow}\partial_x\theta_{i\rho}(x,t)\Ket{R\uparrow}=
    \frac{1+g}{2}\delta_a(x-x_i-v_\rho(t-t_0))
    +\frac{1-g}{2}\delta_a(x-x_i+v_\rho(t-t_0))$
    and a collective spin density excitation 
    $\Bra{R\uparrow}\partial_x\theta_{i\sigma}(x,t)\Ket{R\uparrow}=
    \delta_a(x-x_i-v_\sigma(t-t_0))$, where
    $\delta_a(x)=\frac{1}{\pi}\frac{a}{a^2+x^2}$ \cite{recher02,crepieux03}.
    They propagate with different velocities $v_{\rho,\sigma}$ and have a
    nonzero spatial extent $a$ due to the finite bandwidth. When one of them
    reaches the tunnel point at $x=0$, there is a charge or spin imbalance
    across the junction, which is compensated by a tunneling event: when the
    spin-down excitation in wire~1 arrives at the junction, either a spin-down
    electron can tunnel out of wire~1, or a spin-up electron can tunnel into
    wire~1. So, quite intuitively, spin excitations alone do not create a charge
    current on average \footnote{Spin excitations do, however, create charge
    noise.}.  When, however, the charge excitation in wire~1 arrives at the
    tunnel contact, the charge imbalance induces only tunneling from wire~1 into
    wire~2 [first term in Eq.~(\ref{eq-i2exc})]. It suffices to consider the
    case in which a spin-down electron tunnels \footnote{If a spin-up electron
        tunnels, there cannot be any overlap with the spin-flipped state, as the
        total spin per wire does not vanish.  The event is explicitly ruled out
        by Klein factors in Eq.~\eqref{eq-i2exc}.}. 
    As illustrated in Fig.~\ref{fig-interpretation}(a), an additional charge 
    and an additional spin-down excitation are created in wire~2, and a
    spin-down hole is left behind in wire~1.  This final state is invariant
    under spin flip if the two opposite spin excitations now present in each
    wire compensate. In wire~1, the spin hole must be compensated by the
    spin-down excitation from the injection. Because it is created at the
    position of the \emph{charge} excitation, spin-charge separation makes this
    compensation impossible, unless the injection point in wire~1 is near the
    tunnel junction, so that both excitations are still close. In wire~2, the
    spin-down excitation produced by tunneling needs to coincide with the
    spin-up excitation created at injection, so the process is strong if
    $x_1/v_\rho=x_2/v_\sigma$.  Following the same reasoning, the competing
    process $2\rightarrow 1$ is strongest if $x_2/v_\rho=x_1/v_\sigma$.  Unless
    $v_\rho=v_\sigma$ these conditions cannot be fulfilled simultaneously, so
    the two processes do not cancel and the exchange current becomes nonzero
    [Fig.~\ref{fig-interpretation}(b)]. Using $g v_\rho=v_\sigma$, the two
    conditions can be combined as
    \begin{equation}
        \frac{x_2-x_1}{x_2+x_1}=\pm\frac{g-1}{g+1}
        \label{eq-opti}
    \end{equation}
    and become manifest as extrema in the exchange current (indicated by arrows
    in Fig.~\ref{fig-initial}). Similar peaks reflecting spin-charge separation
    are already present in the direct terms (cf.~Supplemental Material).

    Quantitatively, the final state $1\rightarrow 2$ is
    \begin{equation}
        \Ket{1\rightarrow 2}:=
            \sum_t \psi^\dagger_{2\downarrow}(0,t)\psi_{1\downarrow}(0,t)
                \psi^\dagger_{1\downarrow}(x_1,t_0)
                \psi^\dagger_{2\uparrow}(x_2,t_0)\Ket{},
        \label{eq-estimatestate}
    \end{equation}
    where $t\in\{x_1/v_\rho,x_1/v_\sigma,x_2/v_\sigma\}$ is summed over all
    possible tunnel times (including the spin-induced events to allow for
    interference effects). The corresponding process strength is
    $P_{1\rightarrow 2}:=\braket{\widetilde{1\rightarrow2}|1\rightarrow2}$,
    where $\ket{\widetilde{1\rightarrow2}}$ is the spin-flipped final state,
    obtained by flipping all spin indices in Eq.~\eqref{eq-estimatestate}.
    Constructing $P_{2\rightarrow 1}$ analogously, the exchange current
    becomes
    \begin{equation}
        I_1^\text{(2)exc}\approx e\Gamma_{2e}
            \Bigl|\frac{T}{\hbar v_F}\Bigr|^2 \frac{1+g}{2}
            \Bigl(P_{2\rightarrow 1}-P_{1\rightarrow 2}\Bigr).
        \label{eq-estimate}
    \end{equation}
    All features of the numerics are reproduced by this expression
    (Fig.~\ref{fig-initial}, left inset). From the explicit form of
    $P_{i\rightarrow j}$, which is calculated in the Supplemental Material, we
    extract that the exchange current decays as
    $I^\text{(2)exc}\propto|x_1+x_2|^{-3}$.

    \emph{Biased injection.}---%
    Time controlled pointlike pair injection into QH edge states has recently
    been demonstrated with charge pumps \cite{ubbelohde14}.  A CPS operated at a
    constant voltage, on the other hand, can be modeled by a pair-tunneling
    Hamiltonian which takes into account a voltage induced phase difference
    \cite{mahan81} instead of the initial state approach. The voltage gives rise
    to a length scale $\sim \hbar v_F/eV$. At large bias, $eV\rightarrow
    \hbar v_F a^{-1}$, the injection becomes as pointlike as allowed by
    the bandwidth and we recover the results of the initial state approach. At
    low voltages, it follows from standard renormalization group arguments that
    physical quantities like the current depend on the ratio of $x_{1,2}$ and
    the length scale set by the voltage (Supplemental Material). In particular,
    the exchange currents at two voltages $V$, $V'$ are related through
    \begin{equation}
        I_{1\;x_1,x_2}^\text{exc}(V)
            \approx\Bigl(\frac{V}{V'}\Bigr)^{g^{-1}+g-1}
            I_{1\;\frac{V}{V'}x_1,\frac{V}{V'}x_2}^\text{exc}(V').
        \label{eq-scaling}
    \end{equation}
    This means that in order to access the $x_{1,2}$ dependency illustrated in
    Fig.~\ref{fig-initial} experimentally, it is not necessary to actually move
    the injection points. Rather, varying voltages can be applied to a fixed
    geometry sample (cf.~Supplemental Material). In a QH realization, the length
    of different edges can be fine-tuned by appropriate gating, which
    additionally gives direct access to $x_{1,2}$.
 
    To estimate the signal strength, we assume that the distance between the
    tunnel junction and the injection points is on the order of the length set
    by the bias voltage, $x_1=\hbar v_F/eV$ and $x_2=3\hbar
    v_F/eV$.  At smaller distances injection and tunneling cannot be
    regarded as distinct events. While not changing the physics, this
    considerably complicates quantitative predictions. At much larger distances,
    disorder and spin decoherence may become relevant. Employing the scaling
    relation [Eq.~\eqref{eq-scaling}], we obtain
    $I_{1\;x_1,x_2}^\text{exc}(V)\approx e^2\hbar^{-1}V
    I^\text{exc,init}_{1\;x_1=a,x_2=3a}/(ev_F/a)$, where
    $I_1^\text{exc,init}$ is the exchange current obtained in the initial state
    approach. To remove the explicit dependency on the cutoff we have used
    $g^{-1}+g-1\approx 1$, a valid approximation for a common nanowire
    interaction parameter $g=0.8$. A CPS is operated at voltages below the
    superconducting energy gap of about 1~meV (Nb), such as $V\sim 0.1$~mV. For
    a total transmission $|I|^2|T|^2\sim 10^{-2}\hbar^4 v_F^4$, with $I$
    the injection amplitude, the exchange current is on the order of a few pA, a
    well-accessible value in experiments. With $v_F=10^5$~m/s, the
    injection distances become $|x_1|\sim500$~nm and $|x_2|\sim1500$~nm.

    The primary challenge when designing an entanglement detection scheme which
    is not based on the violation of a Bell-type inequality is to isolate the
    exchange contribution from the background given by the direct contributions
    and measurement noise. This is particularly true for the exchange current,
    which is 2 to 3 orders of magnitude smaller than the background. When
    changing the phase angle $\varphi$ linearly, the exchange contribution
    oscillates and can be isolated easily via lock-in amplification from the
    direct signal, which remains unaffected, cf.~Eq.~\eqref{eq-expvalue}. All
    other parameters can then remain fixed. One way to influence $\varphi$ is
    given by the Rashba spin-orbit interaction \cite{bychkov84} present in
    nanowires: when applying a transversal electric field $E$ (illustrated by
    the gray back gates in Fig.~\ref{fig-setup}), spin-up and spin-down
    electrons acquire different Fermi vectors $k_F\pm k_R$
    \cite{egues02} where $k_R=2\pi/\lambda_R\propto E$ is tunable
    via back gates \cite{liang12,kanai11}.  In this way, until reaching the
    tunnel junction a relative phase
    $\varphi=4\pi(x_1-x_2)/\lambda_\text{R}\propto E$ is collected.  Recent
    experiments on InAs wires show that the Rashba length $\lambda_R$ can
    become as short as 150~nm \cite{wojcik14} which allows for several
    oscillation periods at the beam splitter size as estimated above. When
    ramping the electric field $E$ up and down in a triangular fashion, the
    exchange current oscillates continuously.  Entanglement can thus be detected
    without any magnetic element or correlation measurement.

    To conclude, we have demonstrated how, due to spin-charge separation, the
    hallmark of TLLs, spin entanglement affects the average charge current in an
    electronic beam splitter. The underlying mechanism can be fully understood
    in terms of collective excitations. In addition to traditional entanglement
    detection schemes based on spin filters and correlation measurements, which
    have proven to be notoriously difficult to implement, this effect allows for
    a promising new approach.

    We thank A. Baumgartner, F. Dolcini, T. Fujisawa, and B. Trauzettel for
    helpful discussions and comments.  We acknowledge the support by the EU-FP7
    project SE2ND, No.~271554, the DFG Grant No. RE 2978/1-1 (PR), and Spanish
    MINECO through grant FIS2011-26156 (ALY).

\newcommand{\timeorder}[1][C]{\textrm{T}_\mathcal{#1}}

\setcounter{figure}{0}
\renewcommand{\thefigure}{S\arabic{figure}}

\setcounter{equation}{0}
\renewcommand{\theequation}{S\arabic{equation}}

    \section{Supplemental Material}
        We give further details about the construction of the generating
        functional and the nonequilibrium perturbative expansion within the
        initial state approximation. The direct tunnel contributions to the
        current and to the current noise are discussed with a special focus on
        features induced by spin-charge separation. We give the full analytic
        approximation of the exchange current discussed in the main text and
        confirm the validity of the initial state approximation and the
        bias-dependent scaling behavior numerically. The $I(V)$ and $G(V)$
        behaviors implied by the scaling behavior are plotted. An alternative
        implementation using a quantum Hall sample of Corbino geometry is
        discussed.  For brevity we set $a=e=\hbar=v_F=1$ in intermediate
        results.

    \section{Generating functional and perturbation theory}
        It is convenient to introduce a contour-ordered generating functional
        \begin{align}
            Z^\varphi_{x_1,x_2}&=\Bra{\varphi}\timeorder e^{
                \int dx\int_\mathcal{C}dt\sum_{i\nu s}
                    j_{i\nu s}(x,t)\Phi_{i\nu s}(x,t)}\Ket{\varphi}
                \notag \\
                &=Z^\text{dir}_{x_1,x_2}+\cos(\varphi)\,Z^\text{exc}_{x_1,x_2},
        \end{align}
        where $i\in\{1,2\}$ labels the wire, $\nu\in\{R,L\}\equiv\{1,-1\}$
        distinguishes left and right-movers,
        $s\in\{\uparrow,\downarrow\}\equiv\{1,-1\}$ is the spin index and $j$ a
        source field. Time-ordering $\timeorder$ and the integral in the
        exponential are performed along the Keldysh contour from $t_0$ to
        $\infty$ ($+$~branch) and back again ($-$~branch). The relation between
        the different phase fields reads
        \begin{equation}
            \Phi_{i\nu s}=\Bigl(\phi_{i\rho}+s\phi_{i\sigma}
                +\nu(\theta_{i\rho}+s\theta_{i\sigma})\Bigr).
        \end{equation}
        We can rewrite the current as
        \begin{equation}
            I_1=-\Gamma_{2e}\int dt\sum_{\nu s}\nu\partial_t
                \frac{\delta Z_{x_1,x_2}^\varphi}{\delta j_{1\nu s}(x,t^+)}
                \Bigr|_{j=0},
            \label{eq-i}
        \end{equation}
        and the current noise as
        \begin{align}
            S_{12}=\Gamma_{2e}&\text{Re}\int dt d\tilde t
                \sum_{\nu\tilde\nu s\tilde s}\nu\tilde\nu
                \partial_t\partial_{\tilde t}
                \notag\\
                \Bigl(&
                \frac{\delta^2 Z_{x_1,x_2}^\varphi}{\delta j_{1\nu s}(x,t^-)
                    \delta j_{2\tilde\nu\tilde s}(\tilde x,\tilde t^+)}
                \notag\\
                &\hspace{1cm}
                -\frac{\delta Z_{x_1,x_2}^\varphi}{\delta j_{1\nu s}(x,t^-)}
                \frac{\delta Z_{x_1,x_2}^\varphi}{\delta j_{2\tilde\nu\tilde s}
                    (\tilde x,\tilde t^+)}
                \Bigr)\Bigr|_{j=0}.
            \label{eq-s}
        \end{align}

        Like any expectation value, the generating functional can be written as
        a direct and an exchange term. Explicitly, they read in the interaction
        picture with respect to $H_0$
        \begin{align}
            Z_{x_1,x_2}^\text{dir/exc}=\frac{1}{4}
                &\sum_{\nu_1\nu_1'\nu_2\nu_2'}
                \langle\timeorder 
                F_{1\nu_1\uparrow}F_{2\nu_2\downarrow}
                F^\dagger_{2\nu_2'\downarrow/\uparrow}
                F^\dagger_{1\nu_1'\uparrow/\downarrow} \notag\\
                \exp\Bigl[
                    &\int dx\int_\mathcal{C}dt\sum_{i\nu s}
                        j_{i\nu s}(x,t)\Phi_{i\nu s}(x,t) \notag\\
                    &+2\pi i(\Phi_{1\nu_1\uparrow}(x_1,t_0^-)-
                        \Phi_{1\nu_1'\uparrow/\downarrow}(x_1,t_0^+)) 
                        \notag\\
                    &+2\pi i(\Phi_{2\nu_2\downarrow}(x_2,t_0^-)-
                        \Phi_{2\nu_2'\downarrow/\uparrow}(x_2,t_0^+)) 
                        \notag\\
                    &-i\int_\mathcal{C}dt\ H_T
                \Bigr]\rangle_0
        \end{align}
        with the ground state expectation value $\langle\cdot\rangle_0$ of the
        unperturbed system. To express all quantities of interest we will use
        the contour time-ordered correlation functions, following
        ref.~\cite{delft98},
        \begin{align}
            C^{\nu\nu'}_{ss'}(x,t;x',t')&=
                \langle\timeorder\Phi_{i\nu s}(x,t)
                \Phi_{i\nu's'}(x',t')\rangle \notag\\
                =-\frac{1}{32\pi^2}\Bigl(
                        &(g_\rho^{-1}+\nu+\nu'+\nu\nu'g_\rho)
                            \log(\frac{2\pi}{L}f_{\rho+}) \notag\\
                        +&(g_\rho^{-1}-\nu-\nu'+\nu\nu'g_\rho)
                            \log(\frac{2\pi}{L}f_{\rho-}) \notag\\
                        +&ss'(\rho\rightarrow\sigma)
                    \Bigr),
        \end{align}
        where 
        \begin{equation}
            f_{\alpha\pm}:=-i\text{sgn}_\mathcal{C}(t-t')
                (\pm(x-x')-v_\alpha(t-t'))+a.
        \end{equation}
        The contour sign function $\text{sgn}_\mathcal{C}(t-t')$ is $1$ whenever
        $t$ is later on the Keldysh contour than $t'$, and $-1$ otherwise.

        When expanding $Z$ in the tunnel amplitude, the
        additional Klein factors from the tunnel Hamiltonian impose strong
        constraints on the internal quantum numbers of the tunnel events.
        Using the Debye-Waller identity $\langle e^{\sum x_i}
        \rangle=e^{\frac{1}{2}\langle(\sum x_i)^2\rangle}$ we arrive at
        \begin{widetext}
            \begin{align}
                Z_{x_1,x_2}^{(0)\text{dir}}=\frac{1}{4}\sum_{\nu_1\nu_2}
                    \exp\Bigl[
                    &\frac{1}{2}\int dx dx'
                    \int_\mathcal{C}dt dt'
                    \sum_{i\nu\nu'ss'}
                    j_{i\nu s}(x,t)j_{i\nu's'}(x',t')
                    C^{\nu\nu'}_{ss'}(x,t;x',t') \notag\\
                    &+2\pi i\int dx
                        \int_\mathcal{C} dt\sum_{\nu s}
                        \Bigl[j_{1\nu s}(x,t)\Bigl(
                        C^{\nu\nu_1}_{s\uparrow}(x,t;x_1,t_0^-)
                        -C^{\nu\nu_1}_{s\uparrow}(x,t;x_1,t_0^+)\Bigl) 
                        \notag\\
                        &\hspace{3cm}+j_{2\nu s}(x,t)\Bigl(
                        C^{\nu\nu_2}_{s\downarrow}(x,t;x_2,t_0^-)
                        -C^{\nu\nu_2}_{s\downarrow}(x,t;x_2,t_0^+)
                        \Bigl)\Bigr]
                    \Bigr],
                \label{eq-z0dir}
            \end{align}
            \begin{align}
                Z_{x_1,x_2}^{(2)\text{dir}}=-|T|^2           
                    \int_\mathcal{C}dt' dt''
                    \sum_{\nu_1\nu_2}
                    \exp\Bigl[
                    &\frac{1}{2}\int dx dx'
                    \int_\mathcal{C}dt dt'
                    \sum_{i\nu\nu'ss'}
                    j_{i\nu s}(x,t)j_{i\nu's'}(x',t')
                    C^{\nu\nu'}_{ss'}(x,t;x',t')\Bigr] \notag\\
                \times\Bigl[
                \sum_{\nu'\nu''s'}\exp\Bigl[
                    &+2\pi i\int dx\int_\mathcal{C}dt
                    \sum_{\nu s}
                    \Bigl[j_{1\nu s}(x,t)\Bigl(
                        C^{\nu\nu_1}_{s\uparrow}(x,t;x_1,t_0^-)
                        -C^{\nu\nu_1}_{s\uparrow}(x,t;x_1,t_0^+) \notag\\
                        &\hspace{4cm}+C^{\nu\nu'}_{ss'}(x,t;0,t')
                        -C^{\nu\nu'}_{ss'}(x,t;0,t'')\Bigl) \notag\\
                        &\hspace{3cm}+j_{2\nu s}(x,t)\Bigl(
                        C^{\nu\nu_2}_{s\downarrow}(x,t;x_2,t_0^-)
                        -C^{\nu\nu_2}_{s\downarrow}(x,t;x_2,t_0^+)\notag\\
                        &\hspace{4cm}-C^{\nu\nu''}_{ss'}(x,t;0,t')
                        +C^{\nu\nu''}_{ss'}(x,t;0,t'')\Bigl)\Bigr]
                    \Bigr] \notag\\
                    \times&\Bra{\nu_1\uparrow,\nu_2\downarrow}\timeorder
                        \mathcal{H}^{Ts'}_{1\nu'\rightarrow 2\nu''
                            }(t')
                        \mathcal{H}^{Ts'}_{2\nu''\rightarrow 1\nu'
                            }(t'')
                        \Ket{\nu_1\uparrow,\nu_2\downarrow}
                    \notag\\
                +\sum_{\nu'}\exp\Bigl[
                    &+2\pi i\int dx\int_\mathcal{C}dt
                    \sum_{\nu s}
                    \Bigl[j_{1\nu s}(x,t)\Bigl(
                        C^{\nu\nu_1}_{s\uparrow}(x,t;x_1,t_0^-)
                        -C^{\nu,-\nu_1}_{s\uparrow}(x,t;x_1,t_0^+) \notag\\
                        &\hspace{4cm}+C^{\nu,-\nu_1}_{ss'}(x,t;0,t')
                        -C^{\nu,-\nu_1}_{ss'}(x,t;0,t'')\Bigl)\notag\\
                        &\hspace{3cm}+j_{2\nu s}(x,t)\Bigl(
                        C^{\nu\nu_2}_{s\downarrow}(x,t;x_2,t_0^-)
                        -C^{\nu\nu_2}_{s\downarrow}(x,t;x_2,t_0^+)\notag\\
                        &\hspace{4cm}-C^{\nu\nu'}_{ss'}(x,t;0,t')
                        +C^{\nu\nu'}_{ss'}(x,t;0,t'')\Bigl)\Bigr]
                    \Bigr] \notag\\
                    \times&\Bra{\nu_1\uparrow,\nu_2\downarrow}\timeorder
                        \mathcal{H}^{Ts'}_{1,-\nu_1\rightarrow
                            2\nu'}(t')
                        \mathcal{H}^{Ts'}_{2\nu'\rightarrow 1\nu_1
                            }(t'')
                        \Ket{-\nu_1\uparrow,\nu_2\downarrow} \notag\\
                +\sum_{\nu'}\exp\Bigl[
                    &+2\pi i\int dx\int_\mathcal{C}dt
                    \sum_{\nu s}
                    \Bigl[j_{1\nu s}(x,t)\Bigl(
                        C^{\nu\nu_1}_{s\uparrow}(x,t;x_1,t_0^-)
                        -C^{\nu\nu_1}_{s\uparrow}(x,t;x_1,t_0^+) \notag\\
                        &\hspace{4cm}+C^{\nu\nu'}_{ss'}(x,t;0,t')
                        -C^{\nu\nu'}_{ss'}(x,t;0,t'')\Bigl) \notag\\
                        &\hspace{3cm}+j_{2\nu s}(x,t)\Bigl(
                        C^{\nu\nu_2}_{s\downarrow}(x,t;x_2,t_0^-)
                        -C^{\nu,-\nu_2}_{s\downarrow}(x,t;x_2,t_0^+)
                            \notag\\
                        &\hspace{4cm}-C^{\nu\nu_2}_{ss'}(x,t;0,t')
                        +C^{\nu,-\nu_2}_{ss'}(x,t;0,t'')\Bigl)\Bigr]
                    \Bigr] \notag\\
                    \times&\Bra{\nu_1\uparrow,\nu_2\downarrow}\timeorder
                        \mathcal{H}^{Ts'}_{1\nu'\rightarrow 2\nu_2
                            }(t')
                        \mathcal{H}^{Ts'}_{2,-\nu_2\rightarrow
                            1\nu'}(t'')
                        \Ket{\nu_1\uparrow,-\nu_2\downarrow}\Bigl] \notag\\
            \end{align}
            and
            \begin{align}
                Z_{x_1,x_2}^{(2)\text{exc}}=-|T|^2
                    \int_\mathcal{C}dt' dt''
                    \sum_{\nu_1\nu_1'\nu_2\nu_2'}
                    \exp\Bigl[
                    &\frac{1}{2}\int dx dx'
                    \int_\mathcal{C}dt dt'
                    \sum_{i\nu\nu'ss'}
                    j_{i\nu s}(x,t)j_{i\nu's'}(x',t')
                    C^{\nu\nu'}_{ss'}(x,t;x',t') \notag\\
                    +&2\pi i\int dx\int_\mathcal{C}dt
                    \sum_{\nu s}
                    \Bigl[j_{1\nu s}(x,t)\Bigl(
                        C^{\nu\nu_1}_{s\uparrow}(x,t;x_1,t_0^-)
                        -C^{\nu\nu_1'}_{s\downarrow}(x,t;x_1,t_0^+)\notag\\
                        &\hspace{4cm}+C^{\nu\nu_1'}_{s\downarrow}(x,t;0,t')
                        -C^{\nu\nu_1}_{s\uparrow}(x,t;0,t'')\Bigl) \notag\\
                        &\hspace{3cm}+j_{2\nu s}(x,t)\Bigl(
                        C^{\nu\nu_2}_{s\downarrow}(x,t;x_2,t_0^-)
                        -C^{\nu\nu_2'}_{s\uparrow}(x,t;x_2,t_0^+) \notag\\
                        &\hspace{4cm}-C^{\nu\nu_2}_{s\downarrow}(x,t;0,t')
                        +C^{\nu\nu_2'}_{s\uparrow}(x,t;0,t'')\Bigl)\Bigr]
                    \Bigr] \notag\\
                    \times&\Bra{\nu_1\uparrow,\nu_2\downarrow}\timeorder
                        \mathcal{H}^{T\downarrow}_{1\nu_1'\rightarrow
                            2\nu_2}(t')
                        \mathcal{H}^{T\uparrow}_{2\nu_2'\rightarrow
                            1\nu_1}(t'')
                        \Ket{\nu_1'\downarrow,\nu_2'\uparrow},
                \label{eq-z2exc}
            \end{align}
            where we have decomposed the tunnel Hamiltonian
            $H_T=T\sum_{i\nu\nu's} \mathcal{H}^{Ts}_{i\nu\rightarrow\bar i\nu'}$
            into the processes in which a $\nu$-moving spin-$s$ electron tunnels
            from wire $i$ into wire $\bar i$, becoming a $\nu'$-mover. The
            relevant amplitudes are
            \begin{align}
                \Bra{\nu_1\uparrow,\nu_2\downarrow}&
                    \timeorder
                    \mathcal{H}^{Ts}_{1\nu\rightarrow 2\nu'}(t')
                    \mathcal{H}^{Ts}_{2\nu'\rightarrow 1\nu}(t'')
                \Ket{\nu_1\uparrow,\nu_2\downarrow}=
                \frac{a^{(g^{-1}+g-2)/2}}{4(2\pi)^2}
                \Bigr[
                    f_\rho^{-\frac{g^{-1}+g}{2}}
                    f_\sigma^{-1}
                \Bigl](0,t';0,t'') \notag\\
                &\times\Xi_{\nu_1\nu s}(x_1,t_0^{-};0,t')
                \Xi_{\nu_1\nu s}(x_1,t_0^{+};0,t'')
                \Xi_{\nu_1\nu s}^{-1}(x_1,t_0^{-};0,t'')
                \Xi_{\nu_1\nu s}^{-1}(x_1,t_0^{+};0,t') \notag\\
                &\times\Xi_{\nu_2\nu'\,-s}^{-1}(x_2,t_0^{-};0,t')
                \Xi_{\nu_2\nu'\,-s}(x_2,t_0^{-};0,t'')
                \Xi_{\nu_2\nu'\,-s}(x_2,t_0^{+};0,t')
                \Xi_{\nu_2\nu'\,-s}^{-1}(x_2,t_0^{+};0,t'')
                \label{eq-ampldir}
            \end{align}
            and
            \begin{align}
                \Bra{R\uparrow,R\downarrow}&
                    \timeorder
                    \mathcal{H}^{T\downarrow}_{1R\rightarrow 2R}(t')
                    \mathcal{H}^{T\uparrow}_{2R\rightarrow 1R}(t'')
                \Ket{R\downarrow,R\uparrow}=
                -\frac{a^{(g^{-1}+g+2)/2}}{4(2\pi)^2}
                \Bigr[
                    f_\rho^{-\frac{g^{-1}+g}{2}}
                    f_\sigma
                \Bigl](0,t';0,t'') \notag\\
                &\times\Xi_{RR\downarrow}(x_1,t_0^{-};0,t')
                \Xi_{RR\downarrow}(x_1,t_0^{+};0,t'')
                \Xi_{RR\downarrow}(x_2,t_0^{-};0,t'')
                \Xi_{RR\downarrow}(x_2,t_0^{+};0,t') \notag\\
                &\times\Xi_{RR\uparrow}^{-1}(x_1,t_0^{-};0,t'')
                \Xi_{RR\uparrow}^{-1}(x_1,t_0^{+};0,t')
                \Xi_{RR\uparrow}^{-1}(x_2,t_0^{-};0,t')
                \Xi_{RR\uparrow}^{-1}(x_2,t_0^{+};0,t'')
                \label{eq-amplexc}
            \end{align}
        \end{widetext}
        with the abbreviation
        \begin{align}
            \Xi_{\nu\nu's}=
                f_{\rho+}^{\frac{g^{-1}+\nu+\nu'+\nu\nu' g}{8}}
                f_{\rho-}^{\frac{g^{-1}-\nu-\nu'+\nu\nu' g}{8}}
                f_{\sigma\nu}^{\frac{1+\nu\nu'}{4s}}.
        \end{align}
        Substituting this into Eqs.~\eqref{eq-i} and \eqref{eq-s}, the
        functional derivatives and the $t$ and $\tilde t$-integral can be
        performed in a straightforward fashion. For the integration we use that
        $x,\tilde x\gg a$. The result is independent of the measurement points
        $x$ and $\tilde x$ and the measurement times $t$ and $\tilde t$,
        because, irrespective of when the electrons are injected and of how far
        they have to travel before being measured, they will be detected
        eventually. More importantly, in the second order corrections only terms
        for which the tunneling times $t'$ and $t''$ lie on different branches
        of the Keldysh contour survive. Identifying the integral over the first
        order tunnel Hamiltonian with the interaction picture time evolution
        operator, $-iT\int_{t_0}^\infty dt\mathcal{H}_{i\nu\rightarrow\bar
        i\nu'}^{Ts}=:U_{i\nu\rightarrow\bar i\nu'}^{(1)s}$, this leads to
        the expressions given in the main text, where all
        $2k_F$-processes have been dropped. This is a good approximation
        (which has been checked numerically), because they always require at
        least one initial left-mover to tunnel. In the presence of interactions
        injected left-movers decay into a large left- and a small right-moving
        charge excitation (charge fractionalisation) and the spin excitation
        moves completely to the left, because there are no interactions in the
        spin sector. In the particular geometry under consideration, where the
        injection point is on the left of the tunnel junction, this means that
        the tunneling of initial left-movers is strongly suppressed, because
        most of the excitation does not reach the junction.  In the spin sector
        this is still valid for a beam splitter adiabatically coupled to Fermi
        liquid leads described by a $g(x)$ model
        \cite{dolcini05,lebedev05,recher06}, whereas for the charge excitations
        additional reflections at the boundary to the Fermi liquid leads would
        appear. However, for generic length of the wires, we do not expect
        qualitative changes of the result.

    \section{Direct contributions}
        The direct tunneling corrections of the current and of the current noise
        are 
        \begin{align}
            \label{eq-i2dir}
            I_1&^{(2)\text{dir}}=-e\Gamma_{2e}
                \sum_{\nu_1\nu_2\nu\nu' s}\frac{1+\nu g}{2} \\
                \times&\Bigl[
                    ||U^{(1)s}_{2\nu'\rightarrow 1\nu}
                    \Ket{\nu_1\uparrow,\nu_2\downarrow}||^2
                -||U^{(1)s}_{1\nu\rightarrow 2\nu'}
                \Ket{\nu_1\uparrow,\nu_2\downarrow}||^2
                \Bigr], \notag
        \end{align}
        \begin{align}
            \label{eq-s2dir}
            S_{12}&^{(2)\text{dir}}=-e^2\Gamma_{2e}
                \sum_{\nu_1\nu_2\nu\nu' s}
                \frac{1+\nu g}{2}\frac{1+\nu' g}{2} \\
                \times&\text{Re}\Bigl[
                ||U^{(1)s}_{2\nu'\rightarrow 1\nu}
                    \Ket{\nu_1\uparrow,\nu_2\downarrow}||^2
                +||U^{(1)s}_{1\nu\rightarrow 2\nu'}
                    \Ket{\nu_1\uparrow,\nu_2\downarrow}||^2
                \Bigr]. \notag
        \end{align}
        We notice that the magnitude of these expressions is set by the standard
        quantum mechanical probability of the final state $||\cdot||^2$. This is
        in contrast to the exchange processes discussed in the main text, in
        which the magnitude is given by spin-flipped amplitudes, underlining
        the role of the exchange contribution as a nonclassical interference
        effect. Note that $U$ only includes the part of the tunnel Hamiltonian
        denoted by its indices and is, therefore, not unitary.

\begin{figure*}
            \includegraphics{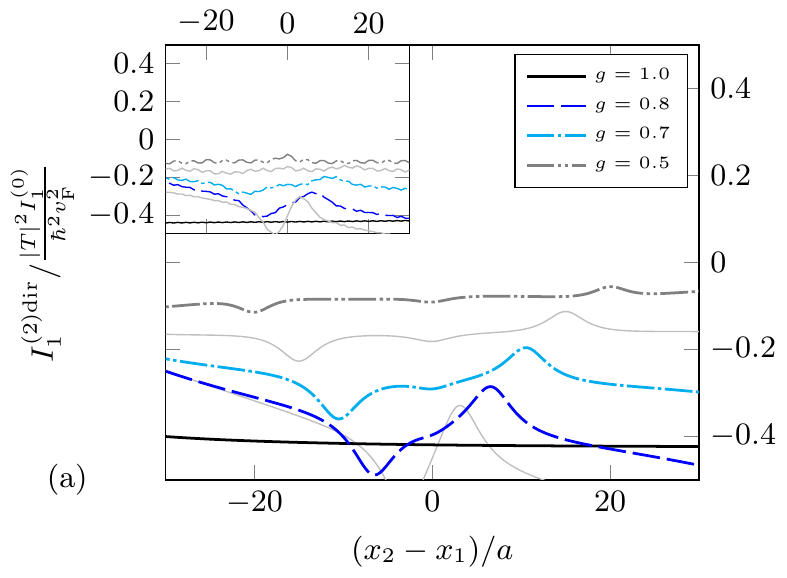}
            \hspace{1cm}
            \includegraphics{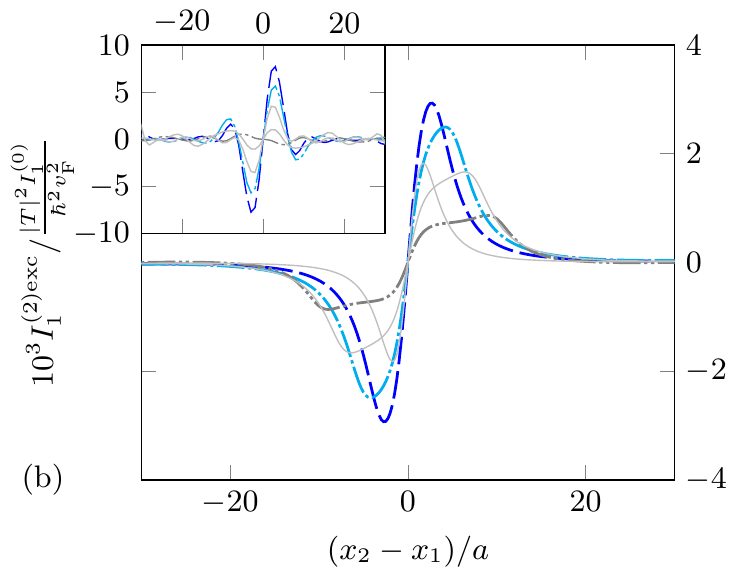}
            
            \includegraphics{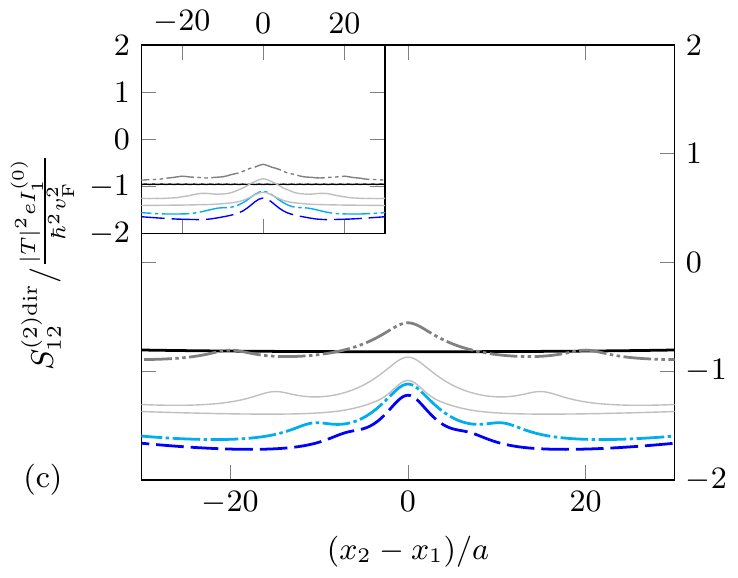}
            \hspace{1cm}
            \includegraphics{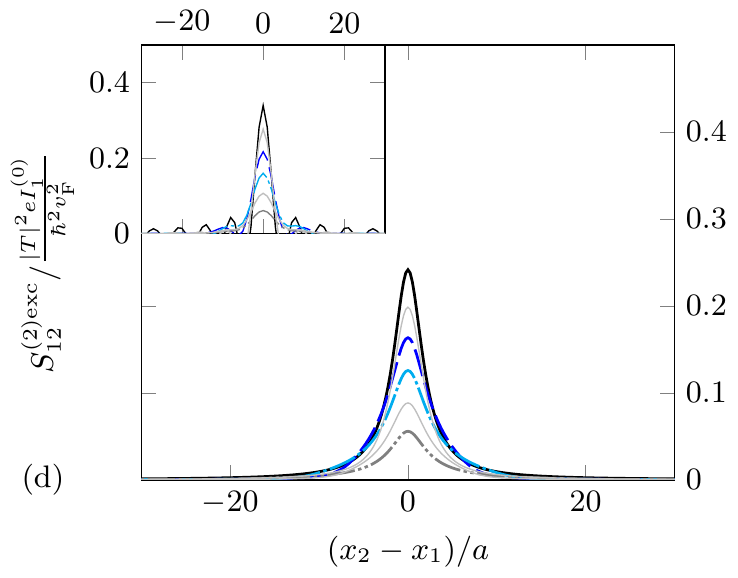}
            \caption{(a) direct and (b) exchange current, (c) direct and (d)
                exchange current noise. Insets: a large voltage bias modeled as
                an injection Hamiltonian with a Peierls phase reproduces the
                initial state approximation for large voltages $2eV\rightarrow
                \hbar v_F a^{-1}$ up to residual oscillations.}
            \label{fig-results}
        \end{figure*}

        The direct contributions are mostly constant, but they show additional
        features for certain injection points $x_{1,2}$
        (Fig.~\ref{fig-results}a,c).
        
        The tunnel correction is always negative, because all tunnel events
        $1\rightarrow 2$ decrease the current after the junction, whereas
        $2\rightarrow 1$ increase it only if the electron becomes a right mover.
        The correction is weaker at higher interaction strength (smaller $g$),
        which is related to the well-known suppression of the tunnel density of
        states in Tomonaga-Luttinger liquids (TLL) \cite{giamarchi03} and can be
        traced back to that fact that electrons are no eigenmodes of the theory.

        In contrast to the exchange current, increasing the distance between the
        injection point and the tunnel junction does not suppress the direct
        current, because spin compensation is not required. As discussed in the
        main text, a charge (spin) imbalance across the junction gives rise to a
        charge (spin) average current. As a consequence there is a tunnel
        current even if the injected electron has completely disintegrated into
        spatially separated charge and spin modes by the time it reaches the
        junction. This holds even though the electron propagator
        $\langle\Psi_\sigma(x,t)\Psi_\sigma^\dagger(0,0)\rangle$ is
        algebraically suppressed in space $x$ and time $t$ (precisely due to
        spin-charge separation), because a Wick decomposition is not allowed in
        this interacting system.

        Similar to the exchange current, there is a two-particle correlation
        effect: tunneling of charge excitations is suppressed if a charge or a
        spin excitation arrives simultaneously at the tunnel junction in the
        other wire.  Two charge excitations can meet only if $x_1=x_2$. In this
        case, both the in- and out-tunneling rates are reduced and so the total
        effect is very small. However, only charge out-tunneling is reduced if
        the charge excitation of electron~1 arrives simultaneously with the
        spin excitation of electron~2, i.e., $x_1/v_\rho=x_2/v_\sigma$.
        Conversely, charge in-tunneling is reduced if the charge excitation of
        electron~2 meets the spin excitation of electron~1 at the junction,
        i.e., $x_1/v_\sigma=x_2/v_\rho$. This produces secondary features, which
        are a signature of spin-charge separation (cf.
        Fig.~\ref{fig-results}a,c).

        The current noise shows the same behavior, except that it is always
        reduced, irrespective of whether in- or out-tunneling is suppressed.
        This changes the sign of half the features, but leaves their position
        unchanged (Fig.~\ref{fig-results}c). The noise is negative, because due
        to charge conservation $\delta I_1$ and $\delta I_2$ have opposite signs
        in all processes.

        Similar side dips have recently been found theoretically in the noise
        of collective excitations in chiral edge channels \cite{wahl14}.

        \begin{widetext}
    \section{Analytic approximation}
        The spin-flipped overlap integral $P_{1\rightarrow 2}$ used in the
        analytic approximation of the exchange current reads
            \begin{align}
                P_{1\rightarrow 2}&:=\sum_{t',t''}
                \langle
                    \psi_{1R\uparrow}(x_1,t_0)
                    \psi_{1R\uparrow}^\dagger(0, t'')
                    \psi_{1R\downarrow}(0, t')
                    \psi_{1R\downarrow}^\dagger(x_1,t_0)
                \rangle_1
                \langle
                    \psi_{2R\downarrow}(x_2,t_0)
                    \psi_{2R\uparrow}(0, t'')
                    \psi_{2R\downarrow}^\dagger(0, t')
                    \psi_{2R\uparrow}^\dagger(x_2,t_0)
                \rangle_2 \notag\\
                &\propto\sum_{t',t''}
                \Bra{R\uparrow,R\downarrow}
                    \timeorder
                    \mathcal{H}^{T\downarrow}_{1R\rightarrow 2R}(t'^{+})
                    \mathcal{H}^{T\uparrow}_{2R\rightarrow 1R}(t''^{-})
                \Ket{R\downarrow,R\uparrow}.
            \end{align}
            Starting from Eq.~\eqref{eq-amplexc}, this can be written as 
            \begin{align}
                P_{1\rightarrow 2}&\propto\sum_{t',t''}
                    \Bigl(\frac{i}{g}(t'-t'')+1\Bigr)^{-\frac{g^{-1}+g}{2}}
                    \Bigl(i(t'-t'')+1\Bigr)
                    \notag\\&\times
                    \Bigl((x_1+t')^2+1\Bigr)^{-\frac{1}{2}}
                    \Bigl((x_1+t'')^2+1\Bigr)^{-\frac{1}{2}}
                    \Bigl((x_2+t')^2+1\Bigr)^{-\frac{1}{2}}
                    \Bigl((x_2+t'')^2+1\Bigr)^{-\frac{1}{2}}
                    \notag\\&\times
                    \exp\Bigl[i\frac{g^{-1}+g+2}{4}
                        \Bigl(
                            \arctan(x_1+t'/g)-\arctan(x_1+t''/g)
                            -\arctan(x_2+t'/g)+\arctan(x_2+t''/g)
                        \Bigr)
                        \notag\\&\hspace{1cm}
                        +i\frac{g^{-1}+g-2}{4}
                        \Bigl(
                            \arctan(-x_1+t'/g)-\arctan(-x_1+t''/g)
                            -\arctan(-x_2+t'/g)+\arctan(-x_2+t''/g)
                        \Bigr)
                    \Bigr],
            \end{align}
        \end{widetext}
        where we chose $t_0=0$. To extract the behavior at large injection
        distances, it suffices to consider absolute values and charge
        fractionalization can be dropped, because it is inessential for the
        spin exchange. So we discard the exponential factor, approximate
        $g^{-1}+g+2\approx4$ and perform a standard Taylor series expansion
        around $(x_1+x_2)^{-1}=0$.

    \section{Biased injection}
        Instead of relying on the initial state approximation, the injection
        can be modeled by a second tunnel Hamiltonian which takes into account
        the voltage induced phase difference as a Peierls phase \cite{mahan81}:
        \begin{align}
            H_I(t)&=Ie^{i\frac{2eV}{\hbar}t}\sum_{\nu_1\nu_2}
                \Bigl[\psi_{1\nu_1\uparrow}(x_1,t)
                    \psi_{2\nu_2\downarrow}(x_2,t)
                    \notag\\&\hspace{2.5cm}
                    +e^{i\varphi}\psi_{1\nu_1\downarrow}(x_1,t)
                    \psi_{2\nu_2\uparrow}(x_2,t)
                \Bigr]+\text{h.c.}
                \notag\\
                &=I\sum_{\nu_1\nu_2\zeta}\Bigl(
                    \mathcal{H}^{I\text{dir}(\zeta)}_{\nu_1\nu_2}
                    +e^{i\varphi}
                    \mathcal{H}^{I\text{exc}(\zeta)}_{\nu_1\nu_2}
                    \Bigr).
            \label{eq-hi}
        \end{align}
        For sufficiently low injection rates there are no correlations between
        subsequently injected pairs and we can restrict the perturbative
        expansion to leading, i.e., second order in the injection amplitude $I$.
        Due to particle number conservation, tunneling and injection each enter
        at even orders only.  All terms which are of zeroth order in the
        injection are equilibrium contributions and do not carry a current or
        cross-wire noise.  The new generating functional is of the form
        \begin{equation}
            Z=Z^{(2I)}+Z^{(2I,2T)}+\dots,
        \end{equation}
        where the superscripts denote the expansion order in injection $I$ and
        tunneling $T$. All expectation values are then taken with respect to the
        ground state. Special care has to be taken when contracting the
        Klein factors. Although not explicitly time dependent they need to be
        time-ordered which can introduce signs. The no-tunneling contributions
        are
        \begin{align}
            I_1^{(2I)}=&-\frac{ev_F}{a}
                \Bigl|\frac{I}{\hbar v_F}\Bigr|^2
                \frac{\pi 2^{4\gamma}}{\Gamma(4\gamma)}
                \text{sgn}(V)
                \Bigl|\frac{ea}{\hbar v_F}V\Bigr|^{4\gamma-1}, \notag\\
        \end{align}
        and
        \begin{align}
            S_{12}^{(2I)}&=\frac{e}{2}|I_1^{(2I)}|,
            \label{eq-i0s0-bias}
        \end{align}
        where $\gamma=\frac{g^{-1}+g+2}{8}$. The current shows a characteristic
        power law behavior which is expected since it applies to any TLL point
        injection calculation \cite{crepieux03,lebedev05}.  Likewise, the
        finite noise contribution is shot noise generated in the injection
        process. Note that the Schottky-like cross-correlation $S_{12}$ does not
        contain an anomalous charge, which has been reported for the
        auto-correlation of infinite TLLs \cite{crepieux03}.

        The leading order contributions are
        \begin{widetext}
            \begin{align}
                I_1^{(2I,2T)\text{dir}}=-|I|^2|T|^2
                    \sum_{\lambda_i\nu_1\nu_2\nu\nu's}
                    &\lambda_0\lambda_1\lambda_2\lambda_3
                    \Bigl(\frac{1+g}{2}(\lambda_0-\lambda_1)
                        +\frac{1+\nu g}{2}(\lambda_2-\lambda_3)\Bigr)
                    \notag \\
                    &\times\int\Bigl(\hspace{-.2cm}\prod_{m=1\dots3}\hspace{-.2cm}d\tau_m\Bigr)
                        \Braket{\timeorder
                            \mathcal{H}^{I\text{dir}}_{\nu_1\nu_2}(0^{\lambda_0})
                            \mathcal{H}^{I\text{dir}\dagger}_{\nu_1\nu_2}
                                (\tau_1^{\lambda_1})
                            \mathcal{H}^{Ts}_{1\rightarrow 2,
                                \nu\rightarrow\nu'}(\tau_2^{\lambda_2})
                            \mathcal{H}^{Ts}_{2\rightarrow 1,
                                \nu'\rightarrow\nu}(\tau_3^{\lambda_3})},
            \end{align}
            \begin{align}
                I_1^{(2I,2T)\text{exc}}=-|I|^2|T|^2\frac{1+g}{2}
                    \sum_{\lambda_i}&\lambda_0\lambda_1\lambda_2\lambda_3
                    \Bigl((\lambda_0-\lambda_1)+(\lambda_2-\lambda_3)\Bigr)
                    \notag \\
                    &\times\int\Bigl(\hspace{-.2cm}\prod_{m=1\dots3}\hspace{-.2cm}d\tau_m\Bigr)
                        \Braket{\timeorder
                            \mathcal{H}^{I\text{dir}}_{RR}(0^{\lambda_0})
                            \mathcal{H}^{I\text{exc}\dagger}_{RR}
                                (\tau_1^{\lambda_1})
                            \mathcal{H}^{T\downarrow}_{1\rightarrow 2,
                                R\rightarrow R}(\tau_2^{\lambda_2})
                            \mathcal{H}^{T\uparrow}_{2\rightarrow 1,
                                R\rightarrow R}(\tau_3^{\lambda_3})},
                \label{eq-bias-i2exc}
            \end{align}
            \begin{align}
                S_{12}^{(2I,2T)\text{dir}}=\frac{|I|^2|T|^2}{2}
                    \sum_{\substack{\lambda_i\nu_1\nu_2\\\nu\nu's}}
                    &\lambda_0\lambda_1\lambda_2\lambda_3
                    \Bigl[\Bigl((\lambda_0-\lambda_1)\frac{1+g}{2}
                            +(\lambda_2-\lambda_3)\frac{1+\nu g}{2}\Bigr)
                        \Bigl((\lambda_0-\lambda_1)\frac{1+g}{2}
                            -(\lambda_2-\lambda_3)\frac{1+\nu'g}{2}\Bigr)\Bigr]
                    \notag \\
                    &\times\int\Bigl(\hspace{-.2cm}\prod_{m=1\dots3}\hspace{-.2cm}d\tau_m\Bigr)
                        \Braket{\timeorder
                            \mathcal{H}^{I\text{dir}}_{\nu_1\nu_2}(0^{\lambda_0})
                            \mathcal{H}^{I\text{dir}\dagger}_{\nu_1\nu_2}
                                (\tau_1^{\lambda_1})
                            \mathcal{H}^{Ts}_{1\rightarrow 2,
                                \nu\rightarrow\nu'}(\tau_2^{\lambda_2})
                            \mathcal{H}^{Ts}_{2\rightarrow 1,
                                \nu'\rightarrow\nu}(\tau_3^{\lambda_3})},
            \end{align}
            \begin{align}
                S_{12}^{(2I,2T)\text{exc}}=\frac{|I|^2|T|^2}{2}
                    \Bigl(\frac{1+g}{2}\Bigr)^2
                    \sum_{\lambda_i}&\lambda_0\lambda_1\lambda_2\lambda_3
                    \Bigl[\Bigl((\lambda_0-\lambda_1)
                            +(\lambda_2-\lambda_3)\Bigr)
                        \Bigl((\lambda_0-\lambda_1)
                            -(\lambda_2-\lambda_3)\Bigr)\Bigr]
                    \notag \\
                    &\times\int\Bigl(\hspace{-.2cm}\prod_{m=1\dots3}\hspace{-.2cm}d\tau_m\Bigr)
                        \Braket{\timeorder
                            \mathcal{H}^{I\text{dir}}_{RR}(0^{\lambda_0}))
                            \mathcal{H}^{I\text{exc}\dagger}_{RR}
                                (\tau_1^{\lambda_1})
                            \mathcal{H}^{T\downarrow}_{1\rightarrow 2,
                                R\rightarrow R}(\tau_2^{\lambda_2})
                            \mathcal{H}^{T\uparrow}_{2\rightarrow 1,
                                R\rightarrow R}(\tau_3^{\lambda_3})},
                \label{eq-bias-s2exc}
            \end{align}
            where $\lambda_i=\pm1$ indicates which part of the Keldysh contour
            the time $\tau^{\lambda_i}$ lies on. At large voltages, a numeric
            integration reproduces the results of the initial state
            approximation (Fig.~\ref{fig-results}, insets).  The overlap
            integrals $\mathcal{I}=\int d\tau\langle\cdots\rangle$ are of
            the general form
            \begin{equation}
                \mathcal{I}(\{x\},V,a)=\frac{1}{a^{\alpha+2}}
                \int\Bigl(\hspace{-.2cm}\prod_{m=1\dots3}\hspace{-.2cm}
                    d\tau_m\Bigr)
                e^{2iV\tau_1}
                \prod_n\Bigl(i(x_n\pm v_n\tau_n)+a\Bigr)^{\alpha_n},
            \end{equation}

            \vspace{1cm}
            up to constant prefactors, where $\alpha=\sum_n \alpha_n$. With
            $\eta$ a number we then derive the exact scaling law
            \vspace{1cm}
            \begin{align}
                \mathcal{I}(\{x\},\eta V,a)&=\frac{1}{a^{\alpha+2}}
                \int\Bigl(\hspace{-.2cm}\prod_{m=1\dots3}\hspace{-.2cm}
                    d\tau_m\Bigr)
                e^{2i\eta V\tau_1}
                \prod_n\Bigl(i(x_n\pm v_n \tau_n)+a\Bigr)^{\alpha_n} \notag\\
                &=\frac{\eta^{-1}}{(\eta a)^{\alpha+2}}
                \int\Bigl(\hspace{-.2cm}\prod_{m=1\dots3}\hspace{-.2cm}
                    d\tilde\tau_m\Bigr)
                e^{2iV\tilde\tau_1}
                \prod_n\Bigl(i(\eta x_n\pm v_n\tilde\tau_n)+
                    \eta a\Bigr)^{\alpha_n} \notag\\
                &=\eta^{-1}\mathcal{I}(\{\eta x\},V,\eta a),
            \end{align}
            where we have substituted $\tilde\tau=\eta\tau$.  Assuming that the
            integral is independent of the high energy cutoff at low
            voltages and/or large injection distances, an approximate scaling
            law can be obtained by neglecting the scaling of $a$ against that
            of $i(x-v\tau)$:
            \begin{align}
                \mathcal{I}(\{x\},\eta V,a)&=\frac{\eta^{-1}}
                    {(\eta a)^{\alpha+2}}
                \int\Bigl(\hspace{-.2cm}\prod_{m=1\dots3}\hspace{-.2cm}
                    d\tilde\tau_m\Bigr)
                e^{2iV\tilde\tau_1}
                \prod_n\Bigl(i(\eta x_n\pm v_n\tilde\tau_n)
                    +\eta a\Bigr)^{\alpha_n} \notag\\
                &\approx\frac{\eta^{-1}}{(\eta a)^{\alpha+2}}
                \int\Bigl(\hspace{-.2cm}\prod_{m=1\dots3}\hspace{-.2cm}
                    d\tilde\tau_m\Bigr)
                e^{2iV\tilde\tau_1}
                \prod_n\Bigl(i(\eta x_n\pm v_n\tilde\tau_n)+a\Bigr)^{\alpha_n} 
                    \notag\\
                &=\eta^{-3-\alpha}\mathcal{I}(\{\eta x\},V,a).
            \end{align}
        \end{widetext}

\begin{figure}
            \includegraphics{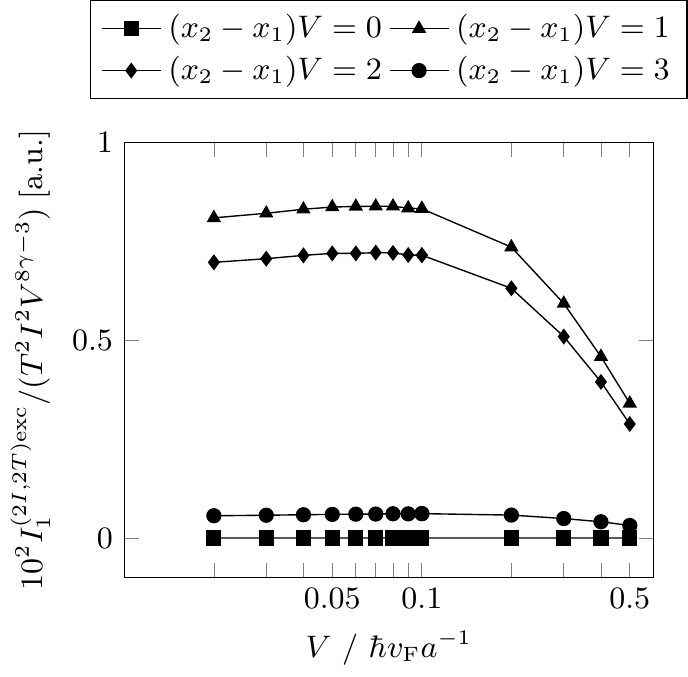}
            \caption{When lowering the voltage and simultaneously rescaling
                $x_1$ and $x_2$, the exchange current follows a power law. At
                large bias voltages, $eV\gtrsim0.1\hbar v_F a^{-1}$, the
                influence of the high energy cutoff is still visible. The
                parameters are $g=0.8$ and $(x_1+x_2)V=-15a$.}
            \label{fig-bias-scaling}
        \end{figure}

        The validity of this approximation has been confirmed numerically
        (Fig.~\ref{fig-bias-scaling}). Specifying $\alpha$ leads to the scaling
        behavior given in the main text.

\begin{figure}
            a)\includegraphics{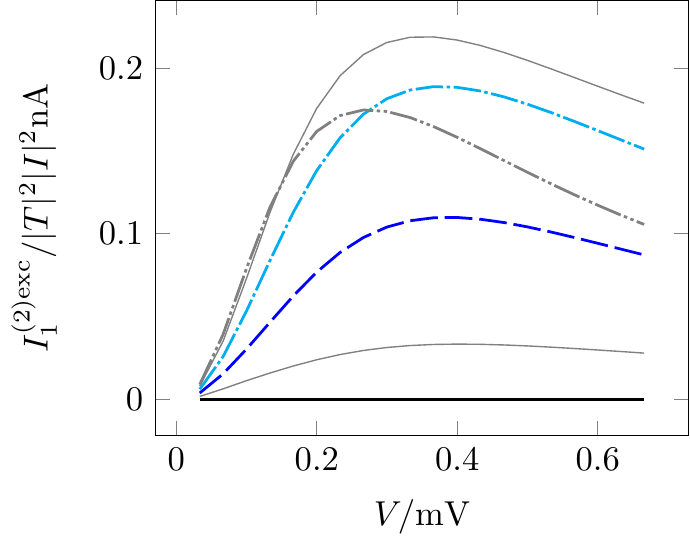}
            \hspace{1cm}
            b)\includegraphics{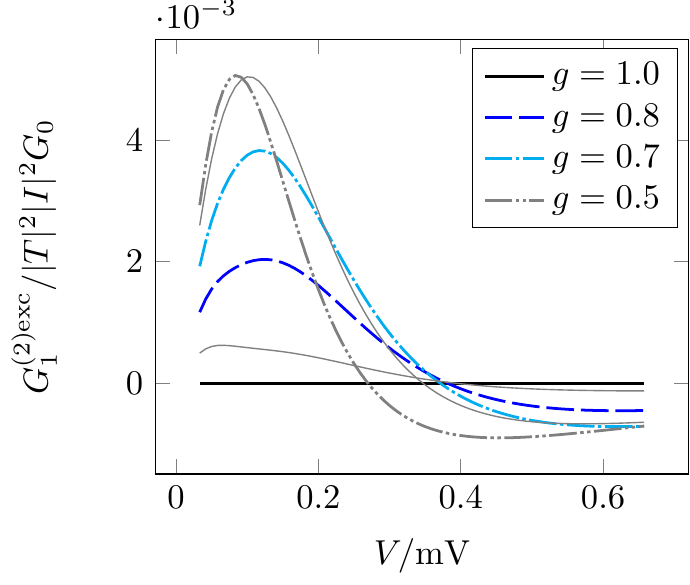}
            \caption{a) Exchange current in wire~1 according to scaling 
                relation (12) for different injection voltages and interaction
                parameters $g$ within an experimentally realistic range. The
                injection points are fixed at $x_1=1500$~nm, $x_2=900$~nm. Gray
                lines represent equidistant intermediate $g$ values. The
                material parameters are the same as in the main text. Note the
                nonmonotonous behavior similar to Fig.~2 in the main text. b)
                The corresponding conductance.}
            \label{fig-bias-current}
        \end{figure}

        The scaling law implies that lowering the injection voltage corresponds
        to increasing all length scales, so the current-voltage relation
        $I(V)$ (Fig.~\ref{fig-bias-current}a) and the conductance-voltage 
        relation $G(V)$ (Fig.~\ref{fig-bias-current}b) pick up the
        nonmonotonous behavior of the exchange current already present in
        Fig.~2. Therefore, it is not necessary in an experiment to actually
        change the injection points $x_{1,2}$.

    \section{Quantum Hall device}        
        Beam splitters have been successfully implemented as integer quantum
        Hall (QH) devices \cite{ji03,neder06,neder07,bocquillon13-1}. Cooper
        pair splitters are likely to suffer from the strong magnetic field, but
        recently a new correlated pair source for QH edge states has been
        realized \cite{ubbelohde14}, which may allow for the injection of
        spin-entangled electrons.  Concerning the beam splitter itself, a QH
        device offers considerable advantages. Being chiral, the edge state of
        the QH effect have a very large mean free path and a coherence length up
        to several 10~$\mu$m. Chirality also implies the absence of charge
        fractionalization, which is an unwanted effect in our device.  The
        intended geometry can be fabricated with high precision and fine tuned
        via appropriate gating in a running experiment thus giving immediate
        access to the injection distances $|x_{1,2}|$. Therefore we will explain
        how the results for the nanowires have to be modified in this section.

        \begin{figure}
            \includegraphics{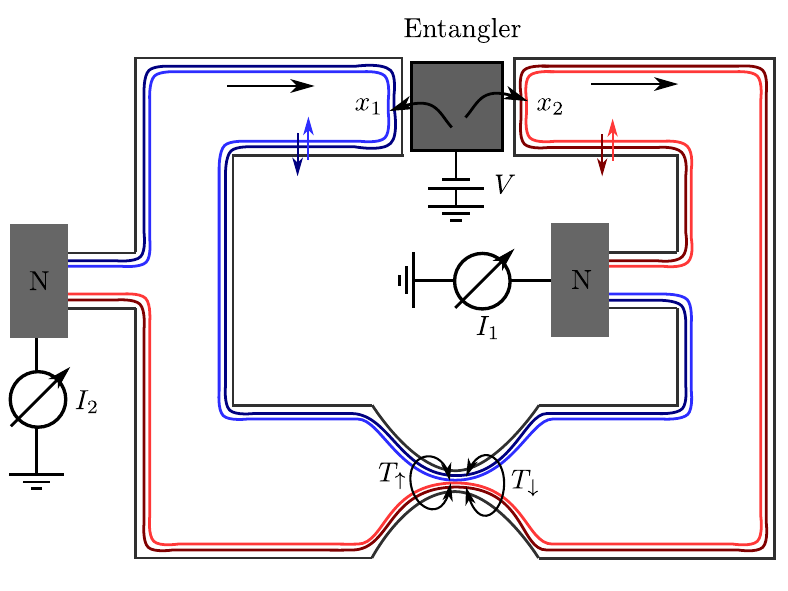}
            \caption{Beam splitter realized in a quantum Hall sample of Corbino
                geometry. Two copropagating channels with opposite spin form a
                spinful chiral TLL at each edge. At a finite Zeeman splitting,
                they are spatially separated and the tunnel amplitudes at the
                constriction become spin dependent.}
            \label{fig-hall}
        \end{figure}

        The beam splitter can be realized as a constriction in a ring-shaped
        Corbino QH sample (Fig.~\ref{fig-hall}). The sample has to be operated
        at a Zeeman splitting much smaller than the Landau splitting, such that
        at each edge two copropagating channels of opposite spin form. Due to
        the difference in Zeeman energy, at a fixed Fermi energy opposite spins
        have different Fermi vectors and are slightly separated in real space.
        This has already been demonstrated experimentally \cite{karmakar11}.
        When linearizing the spectrum around the Fermi points, where we assume
        the Fermi velocity $v_F$ to be independent of edge and spin, the
        electrostatic repulsion between and within the channels can be treated
        exactly using bosonization. This yields two spinful chiral TLLs,
        \begin{equation}
            H_0=\sum_{i\alpha}\int dx\hbar v_\alpha
                \Bigl(\frac{\partial\phi_{i\alpha}}{\partial x}\Bigr)^2,
        \end{equation}
        where $\alpha\in\{\rho,\sigma\}$ labels the charge and the spin channel
        at edge $i\in\{1,2\}$.  As in the nanowire, the charge velocity exceeds
        the spin velocity, $v_\rho>v_\sigma=v_F$. The electron field is
        $\psi_{js}(x)=(2\pi a)^{-1/2}F_{js}\exp\Bigl[ik_s x+2\pi
        i\Phi_{js}(x)\Bigr]$ where
        $\Phi_{js}=\sqrt{\pi/2}(\phi_{j\rho}+s\phi_{j\sigma})$ with
        $s\in\{\uparrow,\downarrow\}$ the spin index and $k_s$ the
        spin-dependent Fermi vector. The tunnel Hamiltonian becomes
        \begin{align}
            H_T&=\sum_s\Bigl(T_s\psi_{1s}^\dagger(0)\psi_{2s}(0)
                +T_s^\star\psi_{2s}^\dagger(0)\psi_{1s}(0)\Bigr) \notag\\
            &=\sum_{s}\Bigl(T_s\mathcal{H}^{Ts}_{2\rightarrow 1}
                +T_s^\star\mathcal{H}^{Ts}_{1\rightarrow 2}\Bigr),
        \end{align}
        where the tunnel amplitude $T_s$ is in general complex because of the
        magnetic field and spin dependent, because at the constriction tunneling
        happens between the outer and the inner channels respectively (cf.
        Fig.~\ref{fig-hall}). TLL behavior has already been investigated and
        observed in this system \cite{bocquillon13-2,wahl14}.

        The initial state is
        \begin{align}
            \Ket{\varphi}&=\sqrt{2}\pi a\Bigl(
                \psi^\dagger_{2\downarrow}(x_2)
                \psi^\dagger_{1\uparrow}(x_1)
                +e^{i\varphi}
                \psi^\dagger_{2\uparrow}(x_2)
                \psi^\dagger_{1\downarrow}(x_1)
                \Bigr)\Ket{}
                \notag\\
            &:=2^{-1/2}(
                \Ket{\uparrow,\downarrow}+
                e^{i\varphi}\Ket{\downarrow,\uparrow}).
        \end{align}
        Lacking SU(2) spin invariance, the direct and exchange contributions
        have to be redefined as
        $\frac{1}{2}(\Bra{\uparrow,\downarrow}\mathcal{O}
        \Ket{\uparrow,\downarrow}+\Bra{\downarrow,\uparrow}\mathcal{O}
        \Ket{\downarrow,\uparrow})$ and
        $\frac{1}{2}(e^{i\varphi}\Bra{\uparrow,\downarrow}\mathcal{O}
        \Ket{\downarrow,\uparrow}+e^{-i\varphi}\Bra{\downarrow,\uparrow}
        \mathcal{O}\Ket{\uparrow,\downarrow})$.

        The calculation proceeds in analogy to the nanowire model and without
        further approximations one finds
        \begin{align}
            \label{eq-i2dir-hall}
            I_1^{(2)\text{dir}}=&-e\Gamma_{2e}
                \frac{|T_\uparrow|^2+|T_\downarrow|^2}{\hbar^2 v_F^2} \\
                \times\sum_{s}&\Bigl[
                    ||U^{(1)s}_{2\rightarrow 1}
                    \Ket{\uparrow,\downarrow}||^2
                -||U^{(1)s}_{1\rightarrow 2}
                \Ket{\uparrow,\downarrow}||^2
                \Bigr], \notag
        \end{align}
        \begin{align}
            \label{eq-i2exc-hall}
            I_1^{(2)\text{exc}}&=e\Gamma_{2e}
                \frac{\text{Re}(T_\uparrow^\star T_\downarrow
                        e^{i(k_\uparrow-k_\downarrow)(x_1-x_2)+i\varphi})}
                    {\hbar^2 v_F^2} \\
                \times&\Bigl[
                \Bra{\uparrow,\downarrow}
                    U^{(1)\uparrow\dagger}_{1\rightarrow 2}
                    U^{(1)\downarrow}_{1\rightarrow 2}
                    \Ket{\downarrow,\uparrow}
                -\Bra{\uparrow,\downarrow}
                    U^{(1)\downarrow\dagger}_{2\rightarrow 1}
                    U^{(1)\uparrow}_{2\rightarrow 1}
                    \Ket{\downarrow,\uparrow}
                \Bigr], \notag
        \end{align}
        \begin{align}
            \label{eq-s2dir-hall}
            S_{12}^{(2)\text{dir}}=&-e^2\Gamma_{2e}
                \frac{|T_\uparrow|^2+|T_\downarrow|^2}{\hbar^2 v_F^2} \\
                \times\sum_{s}&\Bigl[
                    ||U^{(1)s}_{2\rightarrow 1}
                    \Ket{\uparrow,\downarrow}||^2
                +||U^{(1)s}_{1\rightarrow 2}
                \Ket{\uparrow,\downarrow}||^2
                \Bigr], \notag
        \end{align}
        \begin{align}
            \label{eq-s2exc-hall}
            S_1^{(2)\text{exc}}&=-e\Gamma_{2e}
                \frac{\text{Re}(T_\uparrow^\star T_\downarrow
                        e^{i(k_\uparrow-k_\downarrow)(x_1-x_2)+i\varphi})}
                    {\hbar^2 v_F^2} \\
                \times&\Bigl[
                \Bra{\uparrow,\downarrow}
                    U^{(1)\uparrow\dagger}_{1\rightarrow 2}
                    U^{(1)\downarrow}_{1\rightarrow 2}
                    \Ket{\downarrow,\uparrow}
                +\Bra{\uparrow,\downarrow}
                    U^{(1)\downarrow\dagger}_{2\rightarrow 1}
                    U^{(1)\uparrow}_{2\rightarrow 1}
                    \Ket{\downarrow,\uparrow}
                \Bigr], \notag
        \end{align}
        where
        $U^{(1)s}_{j\rightarrow
        k}=-i\int_{t_0}^{\infty}dt'\;\mathcal{H}^T(t')|^s_{j\rightarrow k}$ and
        the amplitudes are
        \begin{widetext}
            \begin{align}
                \Bra{\uparrow,\downarrow}&
                    \timeorder
                    \mathcal{H}^{Ts}_{1\rightarrow 2}(t')
                    \mathcal{H}^{Ts}_{2\rightarrow 1}(t'')
                \Ket{\nu_1\uparrow,\nu_2\downarrow}=
                \frac{1}{4(2\pi)^2}
                \Bigr[
                    f_\rho^{-1}
                    f_\sigma^{-1}
                \Bigl](0,t';0,t'') \notag\\
                &\times\Xi_\uparrow(x_1,t_0^{-};0,t'')
                \Xi_\uparrow(x_1,t_0^{+};0,t')
                \Xi_\uparrow^{-1}(x_1,t_0^{-};0,t')
                \Xi_\uparrow^{-1}(x_1,t_0^{+};0,t'') \notag\\
                &\times\Xi_\downarrow(x_2,t_0^{-};0,t')
                \Xi_\downarrow(x_2,t_0^{+};0,t'')
                \Xi_\downarrow^{-1}(x_2,t_0^{-};0,t'')
                \Xi_\downarrow^{-1}(x_2,t_0^{+};0,t')
                \label{eq-ampldir-hall}
            \end{align}
            and
            \begin{align}
                \Bra{\uparrow,\downarrow}&
                    \timeorder
                    \mathcal{H}^{T\downarrow}_{1\rightarrow 2}(t')
                    \mathcal{H}^{T\uparrow}_{2\rightarrow 1}(t'')
                \Ket{\downarrow,\uparrow}=
                -\frac{a^2}{4(2\pi)^2}
                \Bigr[
                    f_\rho^{-1}
                    f_\sigma
                \Bigl](0,t';0,t'') \notag\\
                &\times\Xi_\uparrow(x_1,t_0^{-};0,t'')
                \Xi_\uparrow(x_1,t_0^{+};0,t')
                \Xi_\uparrow(x_2,t_0^{-};0,t'')
                \Xi_\uparrow(x_2,t_0^{+};0,t'') \notag\\
                &\times\Xi_\downarrow^{-1}(x_1,t_0^{-};0,t')
                \Xi_\downarrow^{-1}(x_1,t_0^{+};0,t'')
                \Xi_\downarrow^{-1}(x_2,t_0^{-};0,t'')
                \Xi_\downarrow^{-1}(x_2,t_0^{+};0,t').
                \label{eq-amplexc-hall}
            \end{align}
        with
        \begin{equation}
            \Xi_s=f_\rho^{-\frac{1}{2}}f_\sigma^{-\frac{s}{2}}
        \end{equation}
        \end{widetext}

        Up to quantitative corrections due to modified exponents we thus recover
        exactly the same behavior as in the nanowire case except that in the
        absence of backscattering, there are as many tunneling events which
        decrease the current after the junction as those which increase the
        current. So the direct contribution is zero up to the additional
        features which are induced by spin-charge separation as discussed in the
        nanowire case. Because the Fermi vectors for spin-up and spin-down are
        different, the exchange current acquires an oscillating prefactor. This
        is essentially the same mechanism as the one induced by the Rashba
        effect in nanowires, and likewise introduces an additional tunable
        parameter.

\end{document}